\documentclass[fleqn,usenatbib]{mnras}

\usepackage[T1]{fontenc}
\usepackage{ae,aecompl}
\usepackage{xcolor}

\usepackage{graphicx}	
\usepackage{amsmath}	
\usepackage{amssymb}	
\usepackage{subfig}
\title[AGN Wind Impact on Planetary Atmospheres]{On the relative importance of AGN winds for the evolution of exoplanet atmospheres}

\author[Heinz]{Sebastian Heinz$^{1}$\\
$^{1}$Department of Astronomy, University of Wisconsin-Madison, 475 N Charter St, Madison, WI 53726, USA\\
\\
}

\date{Accepted 21-Apr-2022. Received 13-Apr-2022; in original form 04-Mar-2022}

\pubyear{2021}

\begin{document}
\label{firstpage}
\pagerange{\pageref{firstpage}--\pageref{lastpage}}
\maketitle

\begin{abstract}
Recent work investigating the impact of winds and outflows from active galactic nuclei (AGN) on the habitability of exoplanets suggests that such activity could be deleterious for the long-term survival of planetary atmospheres and the habitability of planets subject to such winds. Here, we discuss the relative importance of the effect of AGN winds compared to stellar winds and the effect of the planet's magnetosphere and stellar irradiation and conclude that AGN winds are not likely to play a significant role in the evolution of atmospheric conditions in planets under conditions otherwise favorable for habitability.
\end{abstract}


\section{Introduction}
\label{sec:intro}

The question of habitability of exoplanets is generally linked to the surface and atmospheric conditions of such planets. In particular, a planet's ability to retain an atmosphere over billions of years is considered an important prerequisite for that planet to allow surface life to develop.

A number of effects external to the exoplanet may negatively affect the ability of planets to maintain an atmosphere, such as stellar activity and irradiation by nearby high-energy sources. Effects on habitability that are exogenic to the planet's host star system are generally discussed within the framework of a Galactic habitable zone \citep[e.g.][]{gonzalez:01}, which takes account of aspects like Galactic metal abundance and dynamical stability subject to stellar encounters.

We know that most galaxies contain a central supermassive black hole (SMBH) and that, when this black hole accretes gas, it can release copious amounts of energy, making it what is referred to as an active galactic nucleus (AGN). Some effort has been made to understand the impact the central SMBH in our own galaxy, Sgr A*, may have had on the habitability of exoplanets in the Milky Way and our own Earth, and more broadly, what the effect of SMBHs in other galaxies may be on habitable worlds \citep[e.g.][]{amaro-seoane:19,balbi:17,forbes:18,liu:20}.

These studies have focused on the radiative effects of SMBH accretion. Given that we know that accretion in AGN is almost always accompanied by energetic outflows, either in the form or winds or collimated jets, it is important to ask what their effect may be on an exoplanet's atmosphere. \citet{ambrifi:22} recently presented an initial investigation of how AGN outflows from the center of the host galaxy may impact the atmospheres of exoplanets through heating and chemical processing, thus possibly affecting the extent of the Galactic habitable zone.

In \S\ref{sec:comparison} of this manuscript, we review the impact such outflows may have on a planetary system and compare it with other effects and then discuss the ability of planets to withstand the impact of an AGN-driven outflow. We summarize our findings in \S\ref{sec:conclusions}.

\section{The Relative Importance of AGN Winds in Exoplanet Environments}
\label{sec:comparison}

To investigate the effect of outflows from AGN on planetary atmospheres, \citet{ambrifi:22} present a straight forward energy flux calculation, arguing that, under certain conditions, a sub-Eddington outflow from a black hole like Sgr A* can deposit a sufficient amount of energy to thermally unbind a planet's atmosphere over a Salpeter time (the characteristic timescale for AGN accretion), for exoplanets located closer than about 1 kpc from the center of the Galaxy.

The argument relies on calculating the energy input from the outflow upon the top of the atmosphere, under the assumption that the energy is converted to thermal energy that accumulates in the atmosphere. In such a case, it would indeed be possible to unbind a substantial fraction of an Earth-like planet's atmosphere over a Salpeter time. \citet{ambrifi:22} conclude that, for exoplanets located in galaxies harboring more massive or powerful AGN, the critical radius could encompass the entire galaxy.

This conclusion raises the question of whether such atmospheric heating is the dominant energetic term in the energy balance of a planet's atmosphere, and whether an AGN-driven outflow would indeed be able to impinge on a planet's atmosphere.

While a full end-to-end model of planetary energetics in the presence of an AGN wind is beyond the scope of this note, we would like to point out three effects that compete with and/or modulate the impact of the AGN wind on the exoplanet: The host star's wind, the planet's magnetosphere, and the host star's radiative flux.

\subsection{Stellar winds}
Before any outflow from an AGN can interact with a planet's magnetosphere, it must overcome the dynamic pressure of the stellar wind of the planet's host star. That is, we must consider the interaction of two quasi-spherical winds and find the contact discontinuity between the two. This is a well-known problem in the study in interacting stellar wind binaries \citep[e.g.][]{pittard:06} and, in the simplest possible case, boils down to solving for momentum flux balance. In that case, a bow-shock forms around the object with the weaker wind, similar to the heliopause due to the Sun's motion through the interstellar medium, though with different asymptotic bow shock angles that depend on the relative momentum flux of the two winds. 

Clearly, the AGN wind in the case under consideration here is the stronger wind, leading to the formation of a bow shock around the star under consideration.
The point of closest approach of this shock is referred to as the stand-off distance $R_{\rm s}$. any planet orbiting the star with an orbital separation $a$ larger than $R_{\rm s}$ will interact with the impinging AGN wind for part or all of its.

Thus, when determining the effect of an AGN wind on an exoplanet, we should assess the conditions under which the planet's orbit crosses the asteropause, i.e., under what conditions $a>R_{\rm s}$.

Taking the solar wind as a benchmark, we can estimate the stand-off distance from the star outside of which the AGN wind dominates. Using the more relevant case discussed in \citep{ambrifi:22} and following their notation, the case of a thermally/energy drive AGN wind, and adopting their notation, the thrust $\dot{p}$  of the AGN wind is
\begin{equation}
    \dot{p}_{\rm ed} \sim \frac{2\lambda_{\rm Edd}\dot{E}_{\rm Edd}}{v_{\rm ps}}
\end{equation}
where $v_{\rm ps} \sim 1000\,{\rm km\,s^{-1}}$ is the velocity of the energy driven (post-shock) outflow, $\dot{E}_{\rm Edd}$ is the Eddington luminosity and $\lambda_{\rm Edd}=L/L_{\rm Edd}$ is the Eddington ratio. Assuming a quasi-spherical outflow, the momentum flux at distance $r$ from the Galactic center, is then
\begin{equation}
    P_{\rm ed}=\frac{\dot{p}_{\rm ed}}{4\pi r^2}=\frac{\lambda_{\rm Edd}\dot{E}_{\rm Edd}}{2\pi r^2 v_{\rm ps}}
\end{equation}

To find the standoff distance $R_{\rm s}$ from the star, we balance this with the dynamic pressure of the stellar wind, which we parameterize as follows:
\begin{equation}
    P_{\star}=\frac{\dot{M}_{\star}v_{\star}}{4\pi R_{\rm s}^2}
    \label{eq:wind}
\end{equation}
where $v_{\star} \sim 400\,{\rm km\,s^{-1}}\frac{v_{\rm wind}}{v_{\odot}}$ is the characteristic stellar wind velocity and $\dot{M}_{\star} \sim 2\times 10^{-14}\,M_{\odot}\,{\rm yr^{-1}}\frac{\dot{M}_{\star}}{\dot{M}_{\odot}}$ is the 
mass loss rate \citep[e.g][]{versharen:19} which yields
\begin{equation}
    R_{\rm s}=r\sqrt{\frac{\dot{M}_{\star}v_{\star}v_{\rm ps}}{2\lambda_{\rm Edd}\dot{E}_{\rm Edd}}}
\end{equation}

Setting the stand-off distance equal to $a$, we find the critical galacto-centric radius $r_{\rm c}$ inside which an AGN outflow with the given parameters will be able to overcome the stellar wind and impinge on the planet's atmosphere (in the absence of any other shielding effects):
\begin{eqnarray}
    r_{\rm c,\star} & \leq & a\sqrt{\frac{2\lambda_{\rm Edd}\dot{E}_{\rm Edd}}{\dot{M}_{\star}v_{\star}v_{\rm ps}}} \\ & \sim & 420\,{\rm pc}\frac{a}{1AU}\sqrt{\frac{\lambda_{\rm Edd}}{0.05}\frac{M_{\rm BH}}{M_{\rm SgrA*}}\frac{\dot{M}_{\odot}}{\dot{M}_{\star}}\frac{400\,{\rm km/s}}{v_{\star}}\frac{1000\,{\rm km/s}}{v_{\rm ps}}} \nonumber
\end{eqnarray}

This is, in fact, comparable with the galacto-centric radius derived in \citet{ambrifi:22} inside which they conclude heating from an energy-drive AGN wind could evaporate the planet's atmosphere. 

We see that for planets closer to their host star than 1 AU, and for planets orbiting a star with a stronger wind than the Sun, this critical galacto-centric radius $r_{\rm c}$ moves inward (and vice versa for opposite conditions). Given that a predominant interest in the field focuses on planets in the habitable zone around M-dwarfs, which is closer than 1AU, and which may exhibit stronger winds than the solar wind \citep[e.g.][]{wood:21}, we may conclude that stellar winds can effectively shield most exoplanets in the habitable zone from the direct impact of AGN winds even inside the radius $r_{\rm c}$ derived \citet{ambrifi:22}.

One may then also reasonably conclude that in almost all exoplanet candidates, the effects of a stellar wind far outweigh the effects of AGN outflows even under the most generous assumptions about the AGN wind's properties (namely, that the entire outflow energy can be converted into a spherical wind, maximizing the momentum flux). 

Similarly, we can conclude that the momentum flux from the solar wind at Earth corresponds roughly to the critical threshold at which AGN winds should be able to evaporate a planet's atmosphere as discussed in \citet{ambrifi:22}, since the critical radii $r_{\rm c}$ from \citet{ambrifi:22} and $r_{\rm c,\star}$ derived above roughly correspond.

This raises the obvious question of how the atmospheres of such exoplanets, and indeed, the Earth, might survive the energetic effects of the {\em stellar} winds they are subject to. The obvious answer to the latter question is the dynamical shield an exoplanet may raise to ward off the effects of AGN winds, namely: its magnetosphere.

\subsection{Planetary magnetosphere}

Even in cases where a planet's host star does not produce a wind that shields the habitable zone dynamically from AGN winds (or where the stellar wind itself might impose a threat to the survival of its atmosphere), the planet's magnetosphere will provide an effective shield against any partially ionized flow impinging on it.

The interaction of the solar wind with Earth's magnetic field has, of course, been subject to decades of detailed study \citep[e.g.][and references therein]{hartinger:15}. We can see the obvious effect of a planet's magnetic field in the interaction of Earth's dipole field with the solar wind: Our magnetosphere provides an efficient barrier that keeps the solar wind well away from the top of the atmosphere. To estimate the standoff distance of any impinging wind from the planet, we can once again balance the dynamic pressure of the wind 
\begin{eqnarray}
    P_{\rm wind} & = & \frac{\dot{M}_{\rm w}v_{\rm w}}{4\pi R^2}=\frac{\lambda_{\rm Edd}\dot{E}_{\rm Edd}}{2\pi r^2 v_{\rm ps}} \\ & \sim & \frac{5\times 10^{-9}\,{\rm ergs}}{\,cm^{3}}\frac{\lambda_{\rm Edd}}{0.05}\frac{M_{\rm BH}}{M_{\rm SgrA*}}\left(\frac{1\,{\rm kpc}}{r}\right)^{2}\frac{1000\,{\rm km\,s^{-1}}}{v_{\rm ps}} \nonumber 
\end{eqnarray}
with the magnetic pressure of the planet's field (assumed here for simplicity to be a dipole)
\begin{equation}
    \frac{B^2}{8\pi}\sim 4 \times 10^{-3}\,{\rm ergs\,cm^{-3}}\left(\frac{B_{\rm planet}}{B_{\earth}}\right)^{2}\left(\frac{R_{\earth}}{R}\right)^{6}
\end{equation}
to calculate the characteristic size of the planet's magnetosphere \citep[e.g.][]{reda:22}
\begin{equation}
    R_{\rm mag} \sim 10\,R_{\earth}\left(\frac{B}{B_{\earth}}\frac{r}{1\,{\rm kpc}}\right)^{\frac{1}{3}}\left(\frac{0.05}{\lambda_{\rm Edd}}\frac{M_{\rm SgrA*}}{M_{\rm BH}}\frac{1000\,{\rm km\,s^{-1}}}{v_{\rm ps}}\right)^{\frac{1}{6}}
\end{equation}

The steep dependence of the magnetic pressure of a dipole field on distance from the planet's surface immediately indicates that a typical AGN wind will not be able to overcome the dynamic shielding provided by Earth's magnetosphere: For an AGN wind to compress the magnetosphere down to the surface (and thus ablate or at least heat the dense parts of the atmosphere) would require the planet to be located within a critical galacto-centric radius of 
\begin{equation}
    r_{\rm c,B}\leq 1\,{\rm pc}\frac{B_{\earth}}{B}\sqrt{\frac{\lambda_{\rm Edd}}{0.05}\frac{M_{\rm BH}}{M_{\rm SgrA*}}\frac{1000\,{\rm km\,s^{-1}}}{v_{\rm ps}}}
\end{equation}
That is, for a planet with a magnetic field similar in strength to the Earth's, the effect of an AGN wind on the magnetosphere is going to be negligible, and even for much more powerful AGN winds, a Gauss-scale field will provide an efficient shield, as long as the wind is partially ionized (a reasonable assumption). Even under the pessimistic assumption that the AGN wind is entirely neutral, the necessary partial ionization for the AGN wind to interact with the planet's B-field in the vicinity of the exoplanet can likely be provided by the UV-field of the host star.

It worth pointing out that the dynamic pressure assumed here is maximal, i.e., the wind velocity is substantially sub-relativistic. For a relativistic wind or jet, the momentum flux drops inversely proportionally to $v$, correspondingly weakening the dynamic effect it might have on planetary magnetospheres.

\subsection{The relative importance of stellar and AGN winds compared to radiative heating}

For planets with substantially weaker fields, it is plausible that an AGN wind, and certainly a stellar wind, could affect the atmosphere either through ablation or, potentially, dynamical heating---the coupling mode discussed in \citet{ambrifi:22}. How important can we expect such a heating effect to be? To address this question, it is instructive to compare the energy flux delivered by the AGN wind to the energy flux delivered by the stellar radiation field.

The energy flux delivered by an energy-conserving AGN wind (the maximal wind assumption as discussed above) is simply
\begin{eqnarray}
    \dot{Q} & = & \frac{\lambda_{\rm Edd}\dot{E}_{\rm Edd}}{4\pi R^2} \label{eq:agnflux}\\
    & \sim & 0.2\,{\rm ergs\,cm^{-2}\,s^{-1}}\frac{\lambda_{\rm Edd}}{0.05}\frac{M}{M_{\rm SgrA*}}\left(\frac{1\,{\rm kpc}}{R}\right)^{2}\nonumber
\end{eqnarray}
This should be compared to the irradiation by the host star:
\begin{equation}
    \dot{Q}=\frac{L}{4\pi r^2}=1.4\times 10^{6}\,{\rm ergs\,cm^{-2}\,s^{-1}}\frac{L_{\star}}{L_{\odot}}\left(\frac{1\,{\rm AU}}{r}\right)^2
\end{equation}
Clearly, the energetic impact of AGN winds and, in fact, stellar winds, is going to be negligible to first order compared to the radiative flux delivered to the planet in virtually all cases.

In fact, irradiation by the star likely {\em regulates} the size of the planet's magnetosphere \citep{reda:22}, thus relegating the effect of AGN winds to second order even in their impact of the magnetosphere itself.

If AGN winds were relevant for the long-term thermal balance of planetary atmospheres, with the potential to evaporate them, we should expect stellar irradiation to easily unbind planetary atmospheres. The fact that this is not so simply results from thermal balance and the ability of planetary atmospheres to re-radiate the deposited energy by moderately increasing their effective temperatures. 

In other words, even in the absence of a planetary magnetic field and a stellar wind, the energy deposited into an atmosphere by an AGN wind over a Salpeter time will be re-radiated and result in a minor increase in the atmosphere's scale height.  We can reasonably conclude that such heating will not meaningfully affect the habitability of exoplanets.

\subsection{Cosmic Ray Acceleration by AGN wind-driven shocks}

As mentioned in \S\ref{sec:intro}, the impact of high-energy radiation from AGN on atmospheric chemistry of exoplanets has been investigated in a number of publications. In the same vein, it is worth briefly discussing other effects AGN winds might have on exoplanet atmospheres, beyond the impact of high-energy {\em photons} from the AGN.  In particular, energetic charged particles might alter atmospheric chemistry, e.g., through the destruction of ozone \citep[e.g.][]{andersson:17}. The increasingly active nature of late type stars with decreasing mass has been recognized as one of the leading questions in establishing habitability of exoplanets around M-dwarfs \citep[e.g.][]{hu:22}, and Galactic cosmic rays might similarly affect habitability in the close vicinity of a strong cosmic ray sources \citep[e.g.][]{airapetian:18,thomas:20}. A quantitative investigation of atmospheric chemistry and the effect of AGN-accelerated cosmic rays on the Galactic habitable zone is beyond the scope of this work, but we will briefly comment on the potential importance of AGN winds relative to other sources of energetic charged particles.

It is generally understood that the bulk of the Galactic cosmic rays is generated by supernova remnant shocks \citep{reynolds:08,grenier:15}. Overall, the effects of AGN-wind accelerated cosmic rays on exoplanet atmospheres will be similar to those of energetic particles accelerated by supernova remnants. As such, AGN activity will compete with the already existing modes of cosmic ray acceleration. AGN may substantially increase the flux of cosmic rays over the steady state mean flux if they drive large scale shocks into the interstellar medium. The Fermi bubbles suggest that Sgr A* may indeed have been contributing substantially to the overall Galactic cosmic ray energy density at least in the polar regions of the Galactic halo \citep[e.g.][]{yang:22}.

In the solar system,  the bulk of the cosmic ray flux (up to approximately GeV energies) is efficiently shielded by the solar wind and magnetosphere \citep[e.g.][]{jokipii:71}. In fact, solar charged particles dominate the flux of energetic particles near Earth, though their energies are substantially lower than those of bona fide cosmic rays \citep[e.g.,][]{miroshnichenko:13}.

It is possible that---in the absence of a protective stellar wind and planetary magnetosphere---a vastly increased cosmic ray output by Sgr A* might affect atmospheric chemistry of exoplanets, even beyond the potential impact of the ever-present diffuse cosmic ray flux generated by supernova remnants, at least in the central regions of the Galaxy. One of the most intriguing aspects of this potential impact is the suggested formative influence cosmic rays may have had on determining the chirality of life on Earth \citep{globus:20}. A detailed discussion of cosmic rays on the Galactic habitable zone is beyond the scope of this manuscript but clearly deserving of further study. Similarly, regarding the habitability beyond the Milky Way, the effect of strongly Doppler boosted cosmic rays from withing the beaming cone of blazar jets on exoplanet atmospheric chemistry and early life (beyond generic AGN winds considered here and in \citealt{ambrifi:22}) would make for an interesting topic for further study.

Finally, it is also reasonable to ask whether the magnetospheric bow shock of an AGN wind impacting an exoplanet atmosphere can {\em directly} produce cosmic rays by first order Fermi acceleration \citep{bell:78,blandford:78} that might affect atmospheric chemistry. This effect would compete with particle acceleration driven in the atmosphere by the bow shock of the stellar wind of the host star itself and with the flux of energetic particles carried by that wind.

It is instructive to once again estimate the relative energetic importance of such shocks in the context of the solar wind. The solar wind bow shock is, in principle, strong enough to accelerate particles via first order Fermi acceleration \citep{merka:03}.  The approximate kinetic energy flux of the solar wind can be derived by multiplying eq.~\ref{eq:wind} by $v_{\star}/2$:
\begin{eqnarray}
    \dot{Q}_{\star,w} & = & \frac{\dot{M}_{\star}v_{\star}^2}{8\pi R^2} \label{eq:cosmicray}\\
    & \approx & 0.4\,{\rm ergs\, cm^{-2}}\,s^{-1}\frac{\dot{M}_{\star}}{\dot{M}_{\odot}}\left(\frac{v_{\star}}{400\,{\rm km/s}}\frac{1 AU}{a}\right)^{2} \nonumber
    \label{eq:weflux}
\end{eqnarray}
Dividing this exression by $4\pi\,$Sr to estimate the mean cosmic ray intensity, we see that this is roughly an order of magnitude larger than the unmodulated mean intensity in Galactic cosmic rays (which is of order $J_{\rm CR} \sim 2\times 10^{-3}\,{\rm ergs\,cm^{-2}\,s^{-1}\,Sr^{-1}}$ at 1 GeV; e.g., \citealt{blasi:13}), underscoring the fact that solar modulation dominates the arrival of cosmic rays on Earth and the importance of the solar wind in delivering charged particles to Earth.

The energy flux available for cosmic ray acceleration by an AGN wind was already estimated in eq.~\ref{eq:agnflux}. For a reasonable cosmic ray acceleration efficiency of $\epsilon \sim 10\%$ \citep[e.g.][]{caprioli:14} in strong shocks, both expressions should be multiplied by $\epsilon$ for a rough estimate of the resulting cosmic ray energy flux directly accelerated by the respective wind.

Dividing both expressions by $4\pi$\,Sr, we derive estimates of the mean intensities of
\begin{equation}
    J_{\rm CR,\star} \sim 3\times 10^{-3}\,\frac{\rm ergs}{\rm cm^{2}\cdot s\cdot Sr} \frac{\dot{M}_{\star}}{\dot{M}_{\odot}}\frac{\epsilon}{0.1}\left(\frac{v_{\star}}{400\,km/s}\frac{1\,{\rm AU}}{a}\right)^{2}
    \nonumber
\end{equation}
and
\begin{equation}
    J_{\rm CR,AGN} \sim 2\times 10^{-3}\,\frac{\rm ergs}{\rm cm^{2}\cdot s\cdot Sr} \frac{\lambda_{\rm Edd}}{0.05}\frac{\epsilon}{0.1}\frac{M}{M_{\rm SgrA*}}\left(\frac{1\,{\rm kpc}}{R}\right)^{2}
    \nonumber
\end{equation}
for cosmic rays accelerated by the stellar wind and AGN wind bow shocks in the vicinity of the planet, respectively.

Comparing these expressions and the mean Galactic cosmic ray intensity of order $J_{\rm CR} \sim 2\times 10^{-3}\,{\rm ergs\,cm^{-2}\,s^{-2}\,Sr^{-1}}$, we see that the impact of even {\em powerful} AGN winds would only be relevant within about the inner kpc of the Galaxy. However, given that Earth's atmosphere is stable against the charged particle flux it is subjected to by the solar wind (which is an order of magnitude larger still), it is once again reasonable to conclude that the impact of cosmic rays accelerated by AGN winds is of second order, and that habitability will generally be determined by the charged particle flux from the host star to lowest order.

\section{Conclusions} 
\label{sec:conclusions}
While AGN-driven winds are ubiquitous in galaxies and must have occurred during several epochs during the growth period of central supermassive black holes in typical galaxies and thus likely even during Earth's formative epoch, we argue that their energetic impact on exoplanet atmospheres is negligible compared to the effects of stellar irradiation, and that their dynamical impact can also be considered as a second order effect in planets orbiting main sequence stars with stellar winds similar to the solar wind, and/or in planets exhibiting magnetic fields comparable to (or even significantly weaker than) the Earth's.

\section{Data Availability}
Any data used in this manuscript will be shared on request to the corresponding author.

\section{Acknowledgements}
We would like to thank Rich Townsend for helpful discussions and the referee for a constructive report. We acknowledge funding from NASA Astrophysics Theory Grant NNX17AJ98G.


\begin{thebibliography}{}
\makeatletter
\relax
\def\mn@urlcharsother{\let\do\@makeother \do\$\do\&\do\#\do\^\do\_\do\%\do\~}
\def\mn@doi{\begingroup\mn@urlcharsother \@ifnextchar [ {\mn@doi@}
  {\mn@doi@[]}}
\def\mn@doi@[#1]#2{\def\@tempa{#1}\ifx\@tempa\@empty \href
  {http://dx.doi.org/#2} {doi:#2}\else \href {http://dx.doi.org/#2} {#1}\fi
  \endgroup}
\def\mn@eprint#1#2{\mn@eprint@#1:#2::\@nil}
\def\mn@eprint@arXiv#1{\href {http://arxiv.org/abs/#1} {{\tt arXiv:#1}}}
\def\mn@eprint@dblp#1{\href {http://dblp.uni-trier.de/rec/bibtex/#1.xml}
  {dblp:#1}}
\def\mn@eprint@#1:#2:#3:#4\@nil{\def\@tempa {#1}\def\@tempb {#2}\def\@tempc
  {#3}\ifx \@tempc \@empty \let \@tempc \@tempb \let \@tempb \@tempa \fi \ifx
  \@tempb \@empty \def\@tempb {arXiv}\fi \@ifundefined
  {mn@eprint@\@tempb}{\@tempb:\@tempc}{\expandafter \expandafter \csname
  mn@eprint@\@tempb\endcsname \expandafter{\@tempc}}}

\bibitem[\protect\citeauthoryear{{Airapetian} et~al.,}{{Airapetian}
  et~al.}{2018}]{airapetian:18}
{Airapetian} V.~S.,  et~al., 2018, arXiv e-prints, \href
  {https://ui.adsabs.harvard.edu/abs/2018arXiv180303751A} {p. arXiv:1803.03751}

\bibitem[\protect\citeauthoryear{{Amaro-Seoane} \& {Chen}}{{Amaro-Seoane} \&
  {Chen}}{2019}]{amaro-seoane:19}
{Amaro-Seoane} P.,  {Chen} X.,  2019, \mn@doi [\jcap]
  {10.1088/1475-7516/2019/12/056}, \href
  {https://ui.adsabs.harvard.edu/abs/2019JCAP...12..056A} {2019, 056}

\bibitem[\protect\citeauthoryear{{Ambrifi}, {Balbi}, {Lingam}, {Tombesi}  \&
  {Perlman}}{{Ambrifi} et~al.}{2022}]{ambrifi:22}
{Ambrifi} A.,  {Balbi} A.,  {Lingam} M.,  {Tombesi} F.,   {Perlman} E.,  2022,
  arXiv e-prints, \href {https://ui.adsabs.harvard.edu/abs/2022arXiv220300929A}
  {p. arXiv:2203.00929}

\bibitem[\protect\citeauthoryear{Andersson et~al.,}{Andersson
  et~al.}{2018}]{andersson:17}
Andersson M.~E.,  et~al., 2018, \mn@doi [Journal of Geophysical Research:
  Atmospheres] {https://doi.org/10.1002/2017JD027605}, 123, 607

\bibitem[\protect\citeauthoryear{{Balbi} \& {Tombesi}}{{Balbi} \&
  {Tombesi}}{2017}]{balbi:17}
{Balbi} A.,  {Tombesi} F.,  2017, \mn@doi [Scientific Reports]
  {10.1038/s41598-017-16110-0}, \href
  {https://ui.adsabs.harvard.edu/abs/2017NatSR...716626B} {7, 16626}

\bibitem[\protect\citeauthoryear{{Bell}}{{Bell}}{1978}]{bell:78}
{Bell} A.~R.,  1978, \mn@doi [\mnras] {10.1093/mnras/182.2.147}, \href
  {https://ui.adsabs.harvard.edu/abs/1978MNRAS.182..147B} {182, 147}

\bibitem[\protect\citeauthoryear{{Blandford} \& {Ostriker}}{{Blandford} \&
  {Ostriker}}{1978}]{blandford:78}
{Blandford} R.~D.,  {Ostriker} J.~P.,  1978, \mn@doi [\apjl] {10.1086/182658},
  \href {https://ui.adsabs.harvard.edu/abs/1978ApJ...221L..29B} {221, L29}

\bibitem[\protect\citeauthoryear{{Blasi}}{{Blasi}}{2013}]{blasi:13}
{Blasi} P.,  2013, \mn@doi [\aapr] {10.1007/s00159-013-0070-7}, \href
  {https://ui.adsabs.harvard.edu/abs/2013A&ARv..21...70B} {21, 70}

\bibitem[\protect\citeauthoryear{{Caprioli} \& {Spitkovsky}}{{Caprioli} \&
  {Spitkovsky}}{2014}]{caprioli:14}
{Caprioli} D.,  {Spitkovsky} A.,  2014, \mn@doi [\apj]
  {10.1088/0004-637X/783/2/91}, \href
  {https://ui.adsabs.harvard.edu/abs/2014ApJ...783...91C} {783, 91}

\bibitem[\protect\citeauthoryear{{Forbes} \& {Loeb}}{{Forbes} \&
  {Loeb}}{2018}]{forbes:18}
{Forbes} J.~C.,  {Loeb} A.,  2018, \mn@doi [\mnras] {10.1093/mnras/sty1433},
  \href {https://ui.adsabs.harvard.edu/abs/2018MNRAS.479..171F} {479, 171}

\bibitem[\protect\citeauthoryear{{Globus} \& {Blandford}}{{Globus} \&
  {Blandford}}{2020}]{globus:20}
{Globus} N.,  {Blandford} R.~D.,  2020, \mn@doi [\apjl]
  {10.3847/2041-8213/ab8dc6}, \href
  {https://ui.adsabs.harvard.edu/abs/2020ApJ...895L..11G} {895, L11}

\bibitem[\protect\citeauthoryear{{Gonzalez}, {Brownlee}  \& {Ward}}{{Gonzalez}
  et~al.}{2001}]{gonzalez:01}
{Gonzalez} G.,  {Brownlee} D.,   {Ward} P.,  2001, \mn@doi [\icarus]
  {10.1006/icar.2001.6617}, \href
  {https://ui.adsabs.harvard.edu/abs/2001Icar..152..185G} {152, 185}

\bibitem[\protect\citeauthoryear{{Grenier}, {Black}  \& {Strong}}{{Grenier}
  et~al.}{2015}]{grenier:15}
{Grenier} I.~A.,  {Black} J.~H.,   {Strong} A.~W.,  2015, \mn@doi [\araa]
  {10.1146/annurev-astro-082214-122457}, \href
  {https://ui.adsabs.harvard.edu/abs/2015ARA&A..53..199G} {53, 199}

\bibitem[\protect\citeauthoryear{{Hartinger}, {Plaschke}, {Archer}, {Welling},
  {Moldwin}  \& {Ridley}}{{Hartinger} et~al.}{2015}]{hartinger:15}
{Hartinger} M.~D.,  {Plaschke} F.,  {Archer} M.~O.,  {Welling} D.~T.,
  {Moldwin} M.~B.,   {Ridley} A.,  2015, \mn@doi [\grl] {10.1002/2015GL063623},
  \href {https://ui.adsabs.harvard.edu/abs/2015GeoRL..42.2594H} {42, 2594}

\bibitem[\protect\citeauthoryear{{Hu}, {Airapetian}, {Li}, {Zhank}  \&
  {Jin}}{{Hu} et~al.}{2022}]{hu:22}
{Hu} J.,  {Airapetian} V.~S.,  {Li} G.,  {Zhank} G.,   {Jin} M.,  2022, \mn@doi
  [Science Advances] {10.1126/sciadv.abi9743}, 8, 9743

\bibitem[\protect\citeauthoryear{{Jokipii}}{{Jokipii}}{1971}]{jokipii:71}
{Jokipii} J.~R.,  1971, \mn@doi [Reviews of Geophysics and Space Physics]
  {10.1029/RG009i001p00027}, \href
  {https://ui.adsabs.harvard.edu/abs/1971RvGSP...9...27J} {9, 27}

\bibitem[\protect\citeauthoryear{{Liu}, {Chen}  \& {Du}}{{Liu}
  et~al.}{2020}]{liu:20}
{Liu} C.,  {Chen} X.,   {Du} F.,  2020, \mn@doi [\apj]
  {10.3847/1538-4357/aba758}, \href
  {https://ui.adsabs.harvard.edu/abs/2020ApJ...899...92L} {899, 92}

\bibitem[\protect\citeauthoryear{{Merka}, {Szabo}, {{\v{S}}afr{\'a}nkov{\'a}}
  \& {N{\v{e}}me{\v{c}}ek}}{{Merka} et~al.}{2003}]{merka:03}
{Merka} J.,  {Szabo} A.,  {{\v{S}}afr{\'a}nkov{\'a}} J.,
  {N{\v{e}}me{\v{c}}ek} Z.,  2003, \mn@doi [Journal of Geophysical Research
  (Space Physics)] {10.1029/2002JA009697}, \href
  {https://ui.adsabs.harvard.edu/abs/2003JGRA..108.1269M} {108, 1269}

\bibitem[\protect\citeauthoryear{Miroshnichenko \& Nymmik}{Miroshnichenko \&
  Nymmik}{2014}]{miroshnichenko:13}
Miroshnichenko L.,  Nymmik R.,  2014, \mn@doi [Radiation Measurements]
  {https://doi.org/10.1016/j.radmeas.2013.11.010}, 61, 6

\bibitem[\protect\citeauthoryear{{Pittard} \& {Dougherty}}{{Pittard} \&
  {Dougherty}}{2006}]{pittard:06}
{Pittard} J.~M.,  {Dougherty} S.~M.,  2006, \mn@doi [\mnras]
  {10.1111/j.1365-2966.2006.10888.x}, \href
  {https://ui.adsabs.harvard.edu/abs/2006MNRAS.372..801P} {372, 801}

\bibitem[\protect\citeauthoryear{{Reda} et~al.,}{{Reda} et~al.}{2022}]{reda:22}
{Reda} R.,  et~al., 2022, arXiv e-prints, \href
  {https://ui.adsabs.harvard.edu/abs/2022arXiv220301554R} {p. arXiv:2203.01554}

\bibitem[\protect\citeauthoryear{{Reynolds}}{{Reynolds}}{2008}]{reynolds:08}
{Reynolds} S.~P.,  2008, \mn@doi [\araa]
  {10.1146/annurev.astro.46.060407.145237}, \href
  {https://ui.adsabs.harvard.edu/abs/2008ARA&A..46...89R} {46, 89}

\bibitem[\protect\citeauthoryear{{Thomas} \& {Ratterman}}{{Thomas} \&
  {Ratterman}}{2020}]{thomas:20}
{Thomas} B.~C.,  {Ratterman} C.~L.,  2020, \mn@doi [Physical Review Research]
  {10.1103/PhysRevResearch.2.043076}, \href
  {https://ui.adsabs.harvard.edu/abs/2020PhRvR...2d3076T} {2, 043076}

\bibitem[\protect\citeauthoryear{{Versharen}, {Klein}  \& {Maruca}}{{Versharen}
  et~al.}{2019}]{versharen:19}
{Versharen} D.,  {Klein} K.,   {Maruca} B.,  2019, Living Reviews in Solar
  Physics, 28, 5

\bibitem[\protect\citeauthoryear{Wood et~al.,}{Wood et~al.}{2021}]{wood:21}
Wood B.~E.,  et~al., 2021, \mn@doi [The Astrophysical Journal]
  {10.3847/1538-4357/abfda5}, 915, 37

\bibitem[\protect\citeauthoryear{{Yang}, {Ruszkowski}  \& {Zweibel}}{{Yang}
  et~al.}{2022}]{yang:22}
{Yang} H. Y.~K.,  {Ruszkowski} M.,   {Zweibel} E.~G.,  2022, \mn@doi [Nature
  Astronomy] {10.1038/s41550-022-01618-x}, \href
  {https://ui.adsabs.harvard.edu/abs/2022NatAs.tmp...52Y} {}

\makeatother
\end{thebibliography}
\end{document}